\documentclass[prl,amsmath,amsfonts,amssymb,twocolumn]{revtex4}
\usepackage{bbm}
\usepackage{bm}
\usepackage{graphicx}
\usepackage{color}
\usepackage{dsfont}
\usepackage{amsmath}
\graphicspath{{./},{../../Mathematica/KIC/picdump/}}

\definecolor{orange}{RGB}{255,127,0}
\newcommand{\iu}{\mathrm{i}}
\newcommand{\eu}{\mathrm{e}}
\newcommand{\hbare}{\hbar_{\rm eff}}
\newcommand{\jcut}{j_{\rm cut}}
\newcommand{\teq}{\!=\!}
\newcommand{\Tprim}{T_\gamma^{(P)}}
\newcommand{\Nprim}{N_\gamma^{(P)}}

\newcommand{\ie}{\textit{i.e.}}
\newcommand{\Tr}{\operatorname{Tr}}

\begin{document}
\title{Semiclassical Identification of Periodic Orbits in a Quantum Many-Body System}

\author{Maram Akila, Daniel Waltner, Boris Gutkin, Petr Braun and Thomas Guhr}
\affiliation{Fakult\"at f\"ur Physik, Universit\"at Duisburg-Essen,
  Lotharstra\ss e 1, 47048 Duisburg, Germany}

\begin{abstract}
  While a wealth of results has been obtained for chaos in
  single-particle quantum systems, much less is known about chaos in
  quantum many-body systems. We contribute to recent efforts to make a
  semiclassical analysis of such systems feasible, which is
  nontrivial due to the exponential proliferation of
  orbits with increasing particle number. Employing a recently discovered duality relation, we focus
  on the collective, coherent motion that together with the also
  present incoherent one typically leads to a mixture of regular and
  chaotic dynamics. We investigate a kicked spin chain as an example
  of a presently experimentally and theoretically much studied class
  of systems.
\end{abstract}

\maketitle {\it{Introduction}} --- The first step that later on led to
the field of quantum chaos was arguably the introduction of Random
Matrix Theory (RMT) in the early 50's by Wigner to study statistical
aspects of nuclei, for a review see Ref.~\cite{Haake,Stock,GMW1998}. Subsequently,
RMT was applied to model the statistics of atomic and
molecular spectra. Importantly, these systems are self-bound and
interacting many-body systems. In the following decades the insight
emerged that RMT also applies to a single quantum particle in a
complicated potential. Numerical studies of billiards became popular
to explore the connection between classical dynamics and quantum level
statistics~\cite{MDK1979,CVGG1980,Berry1981,BGS1984}, they led to the
celebrated Bohigas-Giannoni-Schmit (BGS) conjecture stating that the
level statistics of the quantum system should be described by RMT if
the corresponding classical system is fully chaotic.  Classical
periodic orbits (POs) and the Gutzwiller trace formula made a detailed
spectral analysis possible \cite{Haake,Stock,cvitano}, \textit{e.g.} for the Hydrogen atom in a
strong magnetic field~\cite{Wintgen,welge,wintgen}, and also a heuristic
understanding of the BGS conjecture~\cite{Berry1985}.  Furthermore, in
the early 80's, the far-reaching connections between mesoscopic and
quantum chaotic systems were uncovered~\cite{Wegner, Efetov}. In the early 00's,
deeper insights into the structure of POs led to much
stronger arguments
supporting the BGS conjecture~\cite{SieberRichter,Heusler}.

The focus on single-particle systems
which had started around 1980 let parts of the quantum chaos community
almost forget that many-body systems were the objects of interest in
the early days of quantum chaos. Only in recent years, new attempts to
address many-body systems in the present  context were put
forward, in particular many-body
localization~\cite{Altshuler,Basko,Znidaric} also observed in recent experiments \cite{bloch,bloch2}, spreading in
self-bound many-body systems~\cite{Haemmerling,Freese}, a
semiclassical analysis of correlated many-particle paths in Bose
Hubbard chains \cite{EnglI} and a trace formula connecting the
energy levels to the classical many-body orbits
\cite{EnglII,Mueller}, to mention just a few. There are also attempts to study field theories semiclassically \cite{cvian}. The first new aspect, specific to many body systems, 
is the existence of two large parameters --- the number of particles $N$ and the dimension of the Hilbert space controlled by the inverse of the  effective Planck constant $\hbare^{-1}$. 
Therefore, different semiclassical limits exist, see Ref.\ \cite{Osipov}.

The second new feature is the complexity of   many-body dynamics. In particular,
many-body systems show collective motion, not present in
single-particle systems. There are various definitions of
collectivity.  Here, we simply mean a coherent motion of all or of
large groups of particles which can be identified in the classical
phase space as well as in the quantum dynamics. Typically, a
many-body system can exhibit incoherent, \textit{i.e.}
non-collective, motion of its particles, coherent collective motion
and forms of motion in between. Which dynamics emerges depends on the
interaction, the excitation energy and the way how the system is
probed. Importantly, collectivity has a strong impact on the level
statistics as is known from numerous analyses of nuclear
spectra. While incoherent particle motion leads to RMT statistics as
in the famous example of the nuclear data ensemble~\cite{guhr_nr1, guhr_nr2},
collective excitations often show Poisson statistics typical to  integrable
systems, as {\textit{e.g.}} in Ref.\ \cite{Guhr} , see Ref.~\cite{GMW1998}  for a review. In general, due to the 
mixed phase space, the BGS conjecture is not directly applicable to many-body systems.

Owing to the focus on the universal regime, the
aforementioned studies of
Bose-Hubbard chains \cite{EnglI,EnglII,Mueller} did not take this
collectivity into consideration, as only   generic properties of chaotic dynamics were   assumed.
To illuminate the full complexity of the motion in many-body
systems and the importance of collectivity from a semiclassical  viewpoint, we consider a
chain of \(N\) spins. This is a many-body generalization of the
kicked top, often used as a model for single particle chaos
\cite{Haake}. We focus on the short time  regime but consider arbitrary $N$, where the  
 collectivity plays a significant r\^ole. 
We have three main goals: First, we want to
establish a new method for the semiclassical analysis  of kicked many-body systems providing 
understanding of the quantum evolution in terms of classical many-particle orbits.
The huge dimension of the Hilbert space might seem to raise an impenetrable barrier   for the  applicability of  semiclassical tools such as the  Gutzwiller trace formula \cite{uzy}. We demonstrate  that this problem can be circumvented  by a suitable generalization
 of the recently discovered duality relation~\cite{Osipov,Akila} that maps properties of the time
evolution and of the enlargement of the system (by adding further
particles) onto each other.  Second, we wish to demonstrate  the importance of 
 collective motion. It may even dominate the quantum spectrum for large particle numbers. Third, we wish to provide a better understanding of spin chain dynamics as this class of systems is 
 presently in the focus
of theoretical~\cite{Braun,Gessner,Atas,Keating} and
experimental~\cite{Simon,Neill,Kim,smith} research.

{\it{System Considered}} --- We study a chain-like kicked
quantum system of \(N\) spins with nearest neighbor interaction
\cite{Prosen} described by the Hamiltonian
\begin{equation}\label{HQM}
 \hat{H}=\hat{H}_I+\hat{H}_K
 \sum_{T=-\infty}^\infty\delta(t-T)
\end{equation}
with the interaction part $\hat{H}_I$ and the kick part $\hat{H}_K$,
\begin{equation}\label{Hexpli}
 \hat{H}_I=\sum_{n=1}^N\frac{4J\hat{s}_{n+1}^z\hat{s}_{n}^z}{(j+1/2)^2},\,\hspace*{2mm}
 \hat{H}_K=\frac{2}{j+1/2}\sum_{n=1}^N{\bf{b}}\cdot\
 \hat{\mathbf{s}}_n\,,
\end{equation}
where $\hat{\bf{s}}_n=(\hat{s}_{n}^x,\hat{s}_{n}^y,\hat{s}_{n}^z)$ are
the operators for spin $n$ and quantum number $j$. Periodic boundary
conditions, \textit{i.e.}  \(\hat{s}_{N+1}^z\!=\!\hat{s}_{1}^z\), make
the system translation invariant. Moreover, \(J\) is the coupling
constant and \(\mathbf{b}\) a magnetic field, assumed without
loss of generality to have the form ${\bf{b}}=(b^x,0,b^z)$. 
We rescaled the terms in Eq.\ (\ref{Hexpli}) by $j+1/2$ in order to keep them 
bounded for $j\rightarrow\infty$. The kicks act at
discrete integer times $T$. 

The one period evolution (Floquet) operator reads
\begin{equation}
 \hat{U}=\hat{U}_I\hat{U}_K,\,
 \hat{U}_I={\rm e}^{-\iu (j+1/2)\hat{H}_I},\,
 \hat{U}_K={\rm e}^{-\iu(j+1/2)\hat{H}_K},
\end{equation}
where \((j+1/2)^{-1}\) takes on the r\^ole of the Planck constant $\hbar_{\rm eff}$.  We
find the corresponding classical system by replacing
\(\hat{\mathbf{s}}_m\to \sqrt{j(j+1)}\,\mathbf{n}_m\) with a classical
spin unit vector precessing on the Bloch sphere \(\mathbf{n}_m\).  
The time evolution can therefore be interpreted as the action
of two subsequent rotation matrices
\begin{equation}
{\bf n}_m(T\!+\!1)=
R_{\bf z} \big( 4J \chi_m\big)
R_{\bf b}\big(2|\mathbf{b}|\big)
{\bf n}_m(T),
\label{eq:ClassRot}
\end{equation}
first around the magnetic field axis and then around the \(z\) axis
(Ising part) with angle \(4J\chi_m\), \(\chi_m\teq n^z_{m-1}+n^z_{m+1}\).  The
classical system can be cast in Hamiltonian form,
\begin{eqnarray}\label{class}
H({\bf{q}},{\bf{p}})&=&\sum_{n=1}^N\left[4J p_{n+1}p_n+\sum_{T=-\infty}^\infty\delta(t-T)\right.
\nonumber\\ &&\left.\times2\left(b^zp_n+b^x\sqrt{1-p_n^2}\cos q_n\right)\right]\,,
\end{eqnarray}
from which the canonical equations follow.  The $N$-component vectors
$\bf{p}$ and $\bf{q}$ are the conjugate momenta and positions of the $N$ (classical)
spins, respectively. The vectors on the Bloch sphere are given by
\begin{equation}\label{normvec}
 {\bf n}_m=\left(\sqrt{1-p_m^2}\cos q_m,\sqrt{1-p_m^2}\sin q_m, p_m
\right)
\end{equation}
in terms of the canonical variables.

\textit{Periodic Orbit Expansion} --- In~\cite{Waltner} we recently
managed to express the trace of the propagator \(\hat{U}\) to power \(T\) for an
interacting spin system in a Gutzwiller-type-of form valid in the limit $j\rightarrow\infty$
\begin{equation}\label{trafor} {\rm Tr}\,\hat{U}^T\sim\sum_{\gamma(T)}
  A_\gamma{\rm e}^{\iu (j + 1/2) S_\gamma}\,.
\end{equation}
This is a sum over classical POs
\(\gamma\) of duration \(T\) if they are well
isolated.  Here, $S_\gamma\allowbreak=\oint_\gamma {\bf p}\cdot
\mathrm{d}{\bf q}-\int\! H({\bf{q}},{\bf{p}})\, \mathrm{d}t$ is the
classical action and, for a sufficiently isolated orbit,
$A_\gamma=\Tprim{\rm e}^{\iu G_\gamma}/\sqrt{\left|\det
    \left(M_\gamma-\mathds{1}\right)\right|}$ with $\Tprim$ the
primitive period of the orbit, $M_\gamma$ the monodromy matrix and
$G_\gamma$ the Maslov phase of $\gamma$.  The \(2N\) eigenvalues
\(\eu^{\pm\lambda_i}\) of \(M_\gamma\) determine the stability of the
PO $\gamma$. For the Hamiltonian \eqref{class}, most POs  are neither fully stable nor  unstable. 
The factor \(A_\gamma\) is finite if  all \(\lambda_i\neq 0\). If, however, \(\lambda_i\teq
0\) for at least one marginal direction, then  \(A_\gamma\) diverges.  This
happens, for instance, if the system undergoes a bifurcation, see~\cite{cat_schomerus}.

The connection between the classical and the quantum system is revealed by
taking the Fourier transform \(\rho(S)\) of Eq. \eqref{trafor} in $j$. This is
methodically similar to Ref.~\cite{welge, wintgen} and was also used on the
single particle kicked top by \cite{Kus,cat_schomerus}. We find
\begin{eqnarray}\label{lengthspec}
 \label{eq:sFT}
 \rho(S)&=&\frac{1}{j_{\rm cut}}\sum_{j=1}^{j_{\rm cut}}{\rm e}^{-\iu (j+1/2)S}{\rm Tr}\,\hat{U}^T\\&{\stackrel{j_{\rm cut}\to\infty}{\sim}}&\,
 \frac{1}{\jcut}\sum_{\gamma(T)} A_\gamma\,\delta(S-S_\gamma)\,,
\nonumber
\end{eqnarray}
which approximates the action spectrum as it
features peaks of width approximately \(\pi/\jcut\) whose
positions are given by the actions modulo \(2\pi\) of the POs
with length \(T\). The peak heights of isolated orbits are
\(A_\gamma\), independent of \(\jcut\).

\textit{Duality Relation} --- At this point, we have to overcome a
severe problem. To resolve the peaks in \(\rho(S)\) we need to compute
${\rm Tr}\,\hat{U}^T$ for sufficiently large \(\jcut\).  But as its
matrix dimension $(2j+1)^N\times (2j+1)^N$ grows exponentially with
\(N\), a direct calculation of the  spectrum of $\hat{U}$ is impossible, \textit{e.g.}, even the propagator \(\hat{U}^T\) for \(N\teq 19\) spins at
\(j=1\) has a matrix dimension of \(10^9\!\times\! 10^9\). Luckily,
recently developed duality relations~\cite{Osipov,Akila} provide the
solution and make possible, for the first time, a semiclassical analysis
of genuine many-body orbits.  The crucial ingredient is the exact
identity
\begin{equation}\label{dual}
\Tr \hat{U}^T=\Tr {\widetilde{U}}^N\, .
\end{equation}
The trace over the time-evolution operator $\hat{U}$ for $T$ time
steps equals the trace over a ``particle-number-evolution'' operator
$\widetilde{U}$ for $N$ particles. The form of the non-unitary dual
operator $\widetilde{U}$ is similar to that of the time-evolution
operator for a chain of \(T\) spins. Its dimension \((2j+1)^T \times
(2j+1)^T\) is governed by the time $T$ while the dimension $(2j+1)^N\times
(2j+1)^N$ of the operator $\hat{U}$ grows with the particle number
$N$. This duality allows us to calculate \(\rho(S)\) for arbitrary
\(N\) as long as \(T\) is sufficiently short. We generalize this duality approach,
originally developed for $j=1/2$~\cite{Akila}, to
$j\gg1$ to make it applicable in the present context~\cite{AkilaI}. The form of
$\widetilde{U}$ is given in the supplemental material.


\textit{Incoherent versus Collective Motion} --- Anticipating the
results of the subsequent detailed analysis, we sketch the r\^ole of
some collective orbits as opposed to the incoherent ones. Given
translational invariance, a PO  of  the \(M\)-particle system generates a PO  in   the  system with \(k M\)
particles for any integer $k$. We therefore introduce \(\Nprim\),
analogous to \(\Tprim\), as the minimal number of particles required
for a PO  to close in the \(N\)-body system. Such orbits with relatively small
\(\Nprim\teq N/k\) yield a form of collective dynamics
as the spins \(n\) and \(n+\Nprim\) move
synchronously. The action of these orbits scales linearly with \(N\),
\(S_\gamma\teq (N/\Nprim)S_\gamma^P\), where \(S_\gamma^P\) is the
action for \(\Nprim\) particles.  Moreover, if, for example, an orbit
for \(N\) spins has a marginal direction, this implies a marginal direction in the corresponding PO for \(kN\) spins. The number of periodic
orbits grows strongly with \(N\) which limits their  resolution.
However, collective orbits can stick out in such a
remarkable way that they  dominate the spectra. 


\begin{figure}
 \includegraphics[width=0.9\columnwidth]{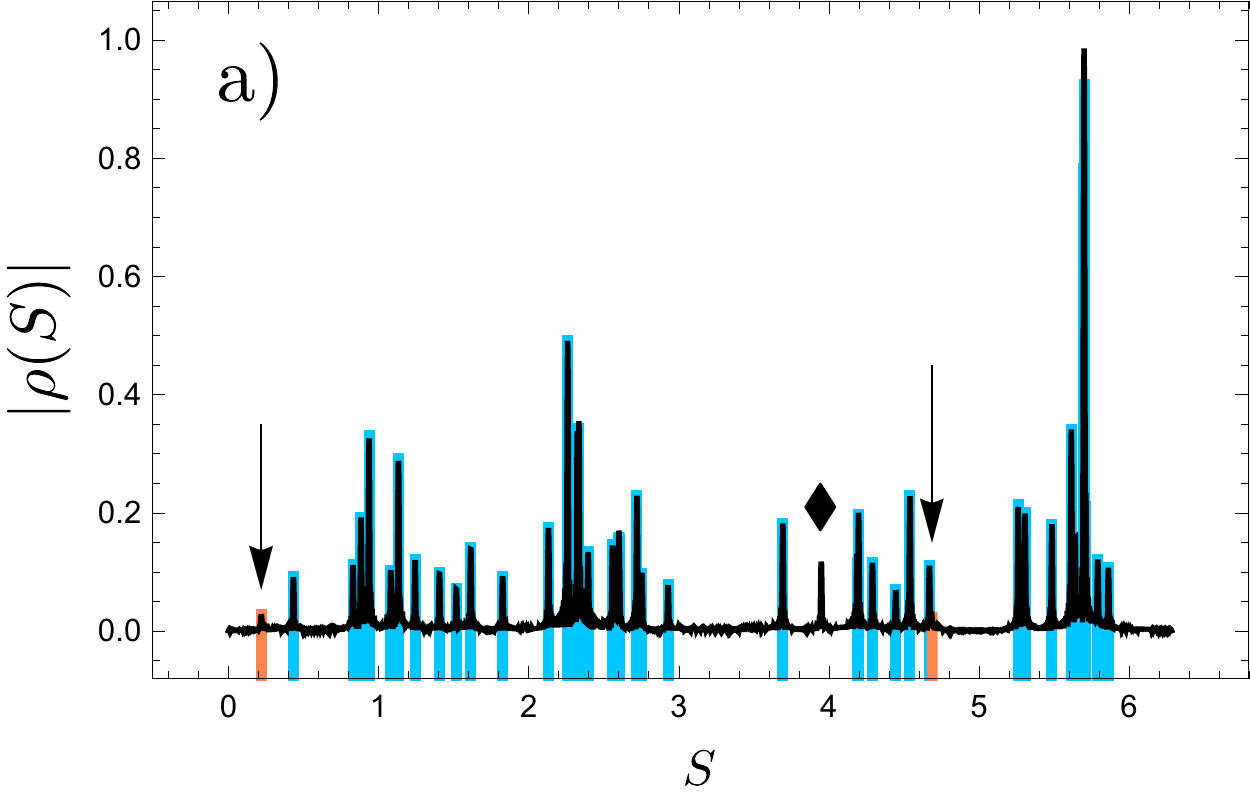}\\
\includegraphics[width=0.9\columnwidth]{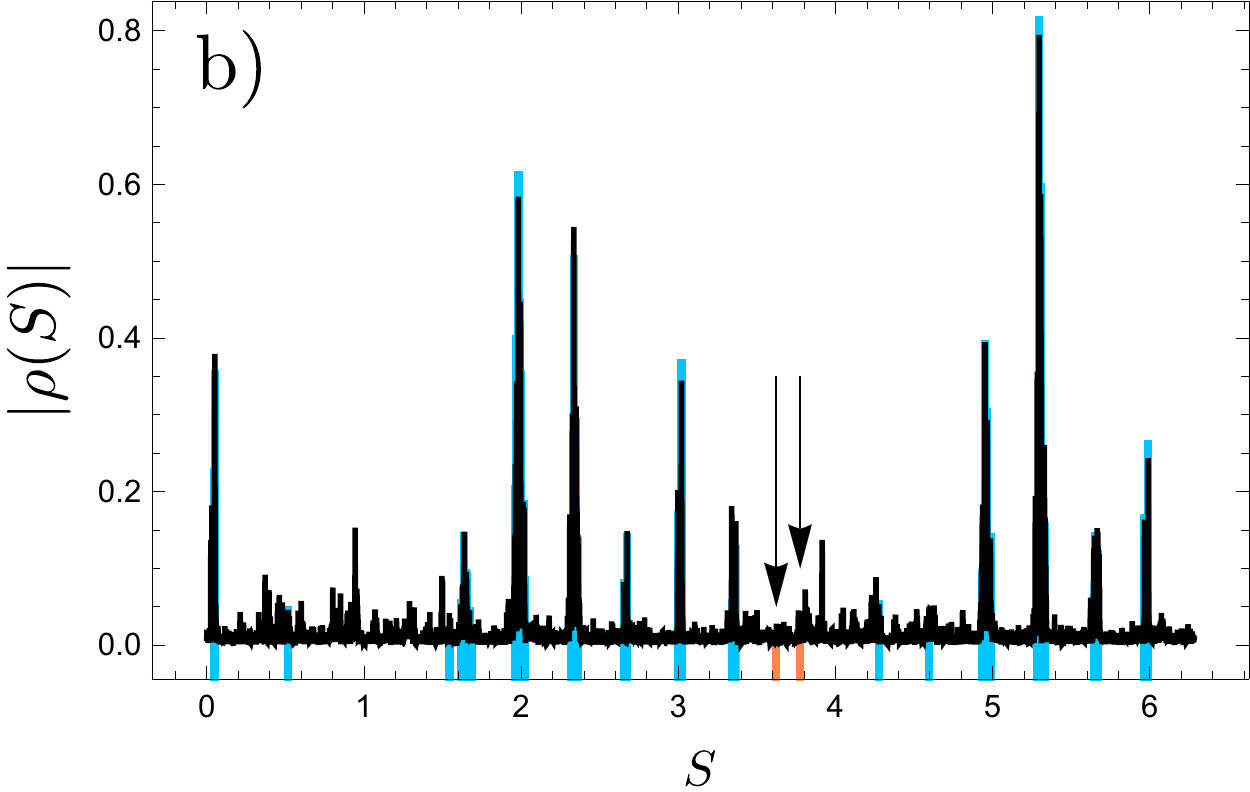}\\
 \caption{For $T=1$, $|\rho(S)|$ as black curve. a) \(N\teq7\) spins with \(J=0.75\) and \(b^x\teq b^z\teq0.9\), cut-off \(\jcut\teq 800\).
 b) $N=19$ spins with $J=0.7$, $j_{\rm cut}=4650$ and ${\bf b}$ as in a).
 The positions of the classical POs as red lines for \(\Nprim\teq1\), as blue lines for \(\Nprim\teq N\).
 }
 \label{fig1}
 \label{fig:sFTT1}
\end{figure}
\textit{Details of the Analysis} --- To demonstrate the power of our
method we provide a numerical calculation of $|\rho(S)|$ for \(T\teq 1\) and \(N\teq 7\)
spins in Fig.\ \ref{fig:sFTT1}(a). In this case the number of POs
is sufficiently low for a semiclassical analysis of individual peaks.
The black peaks result from the Fourier transformation of ${\rm Tr}\,{\hat U}^T$, the positions of the colored peaks 
are given by the actions $S_\gamma$ of the POs derived from the Hamiltonian (\ref{class}).
 Both, collective    \(\Nprim=1\) and non-collective  \(\Nprim=7\) types of POs are present. The former are 
highlighted  by arrows in Fig.\ \ref{fig:sFTT1}(a). The action spectrum  \(S_\gamma\) of  POs  is well reproduced, by the positions of the maxima of \(|\rho(S)|\). The only exception is a peak
marked by \(\blacklozenge\) in Fig.\ \ref{fig:sFTT1}(a), which  is
due to a complex predecessor of a nearly bifurcating orbit. The existence of 
 such ghost orbits is known in the context of   single-particle systems~\cite{Kus}. 

\begin{figure}
\includegraphics[width=0.9\columnwidth]{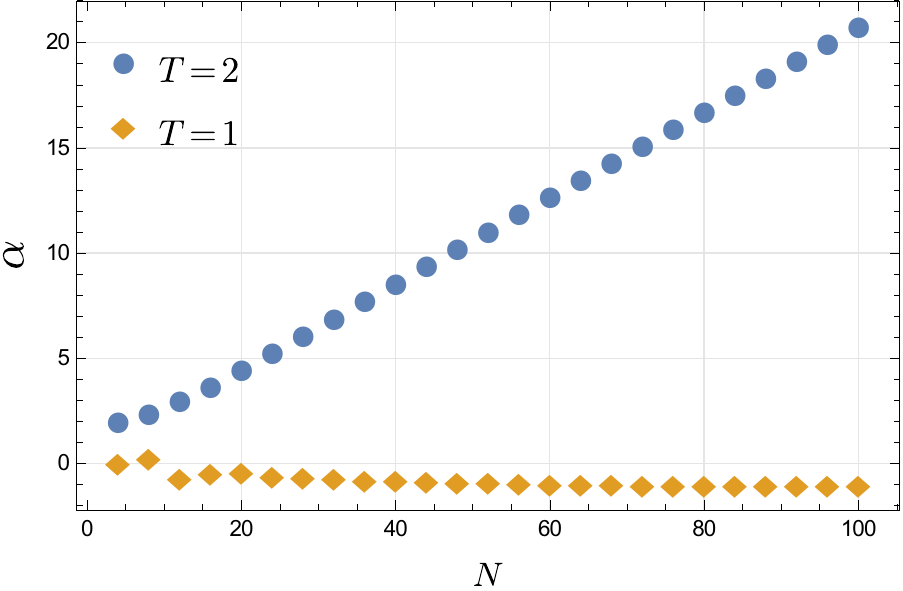}
\caption{Scaling exponent \(\alpha\) of $|\rho(S_\gamma)|\sim(j_{\rm cut})^\alpha$
  at the position \(S_\gamma\) of the largest peak,
versus the number of spins. For \(T\teq 1,2\), \(J\teq 0.7\), \(b^x\teq b^z \teq 0.9\) and \(N\teq 4k\).}
\label{fig:alphascale}
\end{figure}

In
Fig.~\ref{fig:sFTT1}(b) we depict \(|\rho(S)|\), for \(T\teq 1\),
but now for \(N\teq 19\) spins. Since  
the number of POs  grows exponentially with  \(N\) it is not possible in this case  to resolve all of them for computationally feasible
 values of $\jcut$.  Nevertheless, it is possible to identify those that provide the most significant contribution to ${\rm Tr}\,{\hat U}^T$. Therefore we   employ a simple 
filtering criterion by selecting only POs   satisfying  \(A_\gamma^{-2}\propto|\det{(
  M_\gamma-\mathds{1} )} |< 10^6\).  As one can see, the actions of these orbits reproduce the positions of  the most significant spikes of \(|\rho(S)|\). The semiclassical reconstruction of \(|\rho(S)|\) becomes additionally more challenging with growing $N$  due to the increasing number of nearly bifurcating  POs. In contrast to an isolated   PO where  \(|\rho(S)|\) does not scale with $\jcut$, we find for bifurcating orbits a nontrivial scaling  \(|\rho(S_\gamma)|\sim (\jcut)^\alpha\), where 
 \(\alpha\) depends on the bifurcation type.  Importantly, 
for $T=1$ the exponent \(\alpha \) does not grow with $N$, see the lower dotted line in Fig.~\ref{fig:alphascale}.  For a single degree of freedom the effect of bifurcations was studied in \cite{cat_manderfeld,cat_schomerus}. We found
these results consistent with our numerics.

\begin{figure*}[thb]
\hfill
\includegraphics[width=0.32\textwidth]{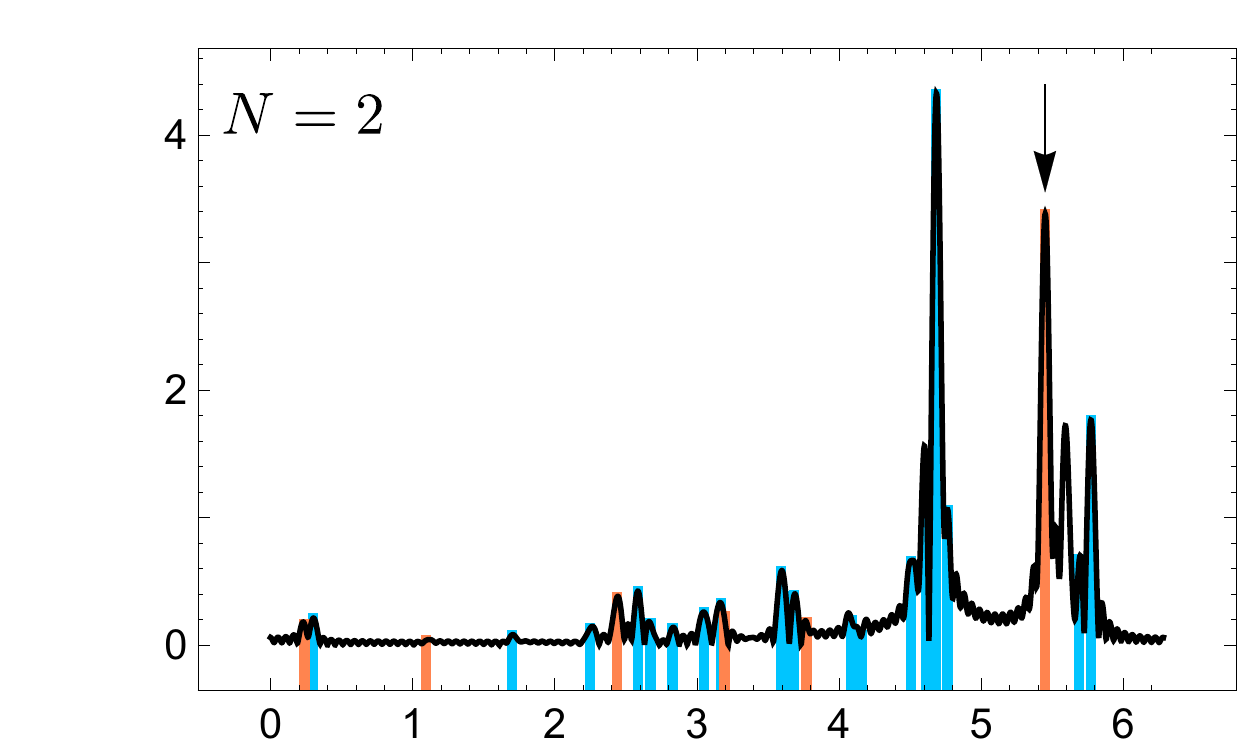}
\hfill
\includegraphics[width=0.32\textwidth]{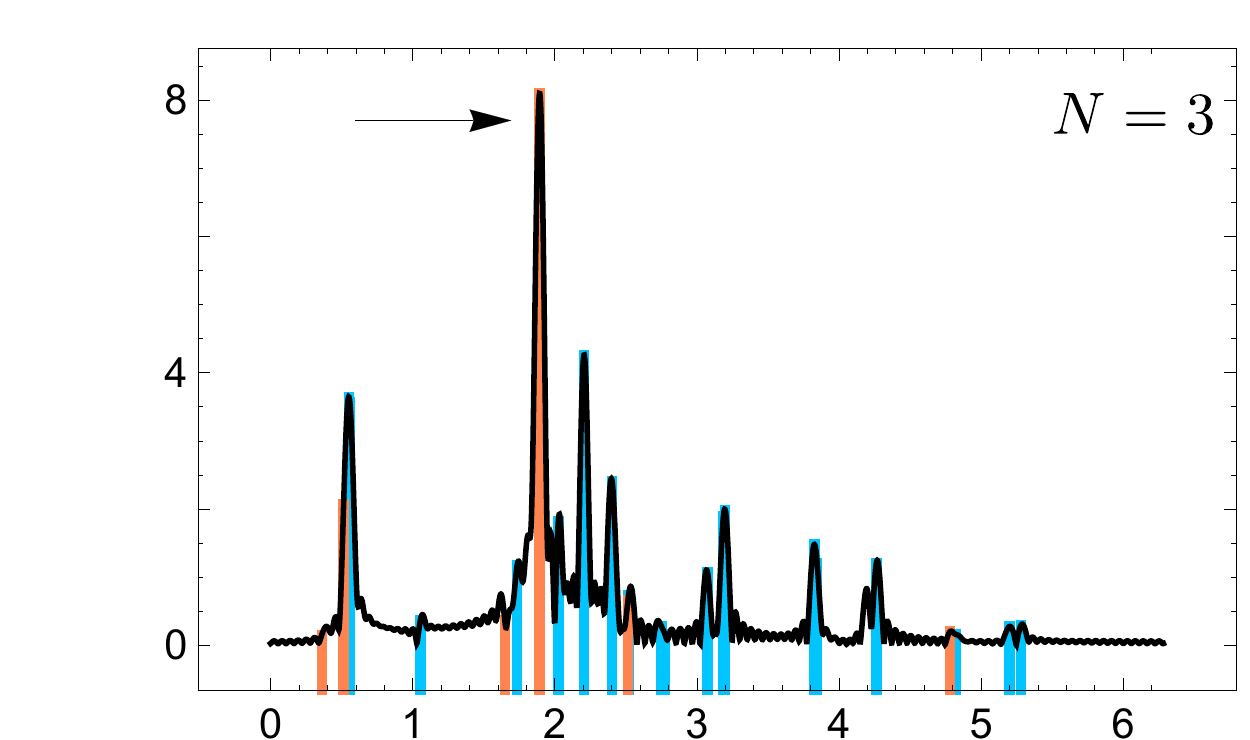}
\hfill
\includegraphics[width=0.32\textwidth]{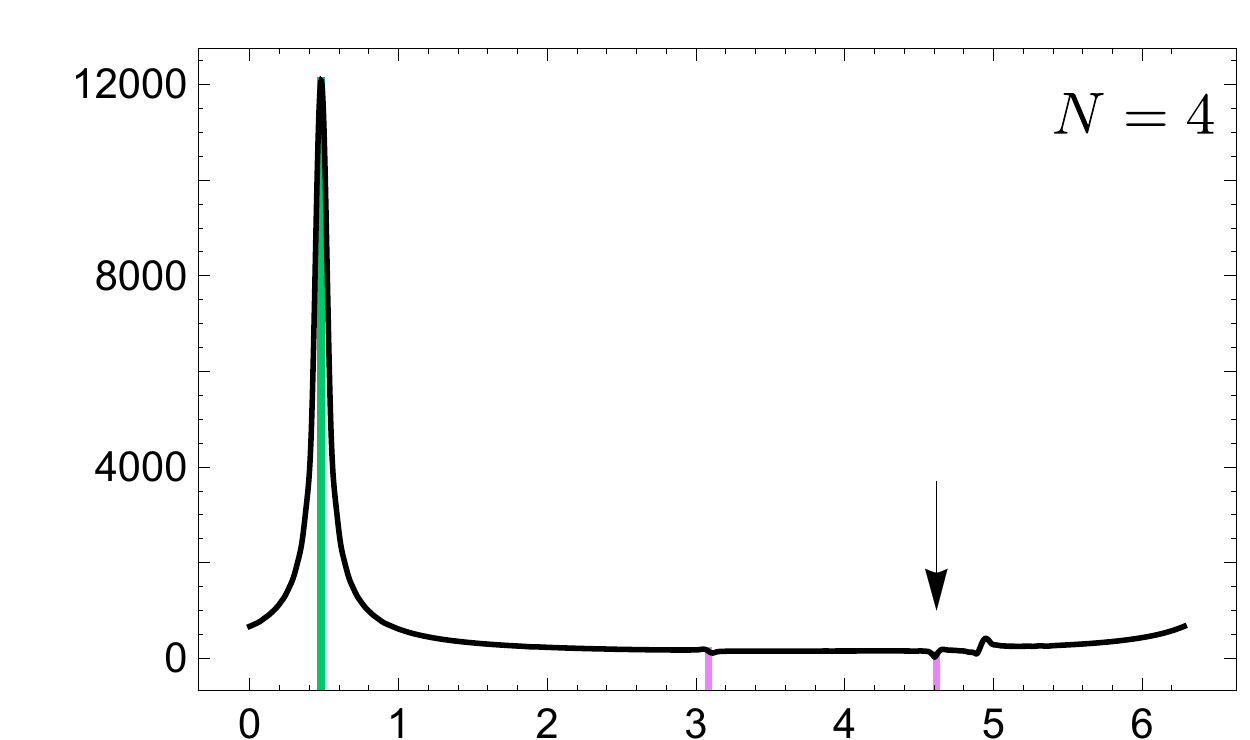}
\\
\hfill
\includegraphics[width=0.32\textwidth]{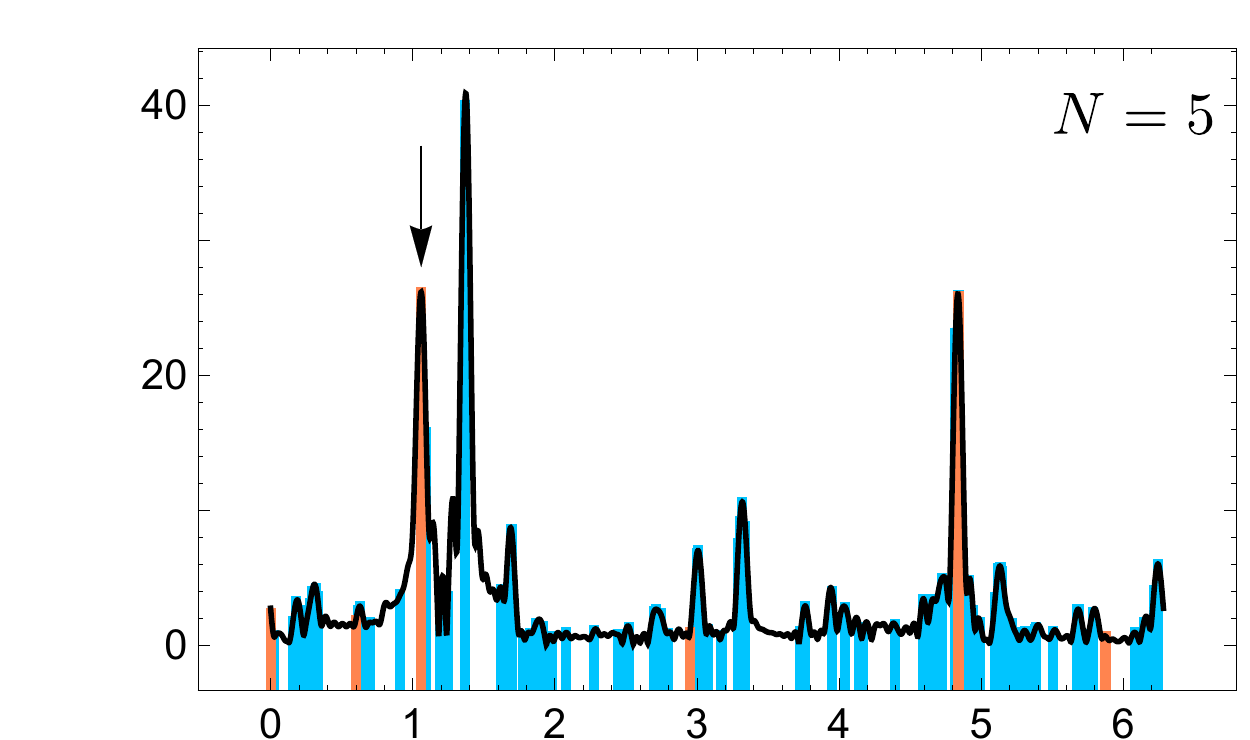}
\hfill
\includegraphics[width=0.32\textwidth]{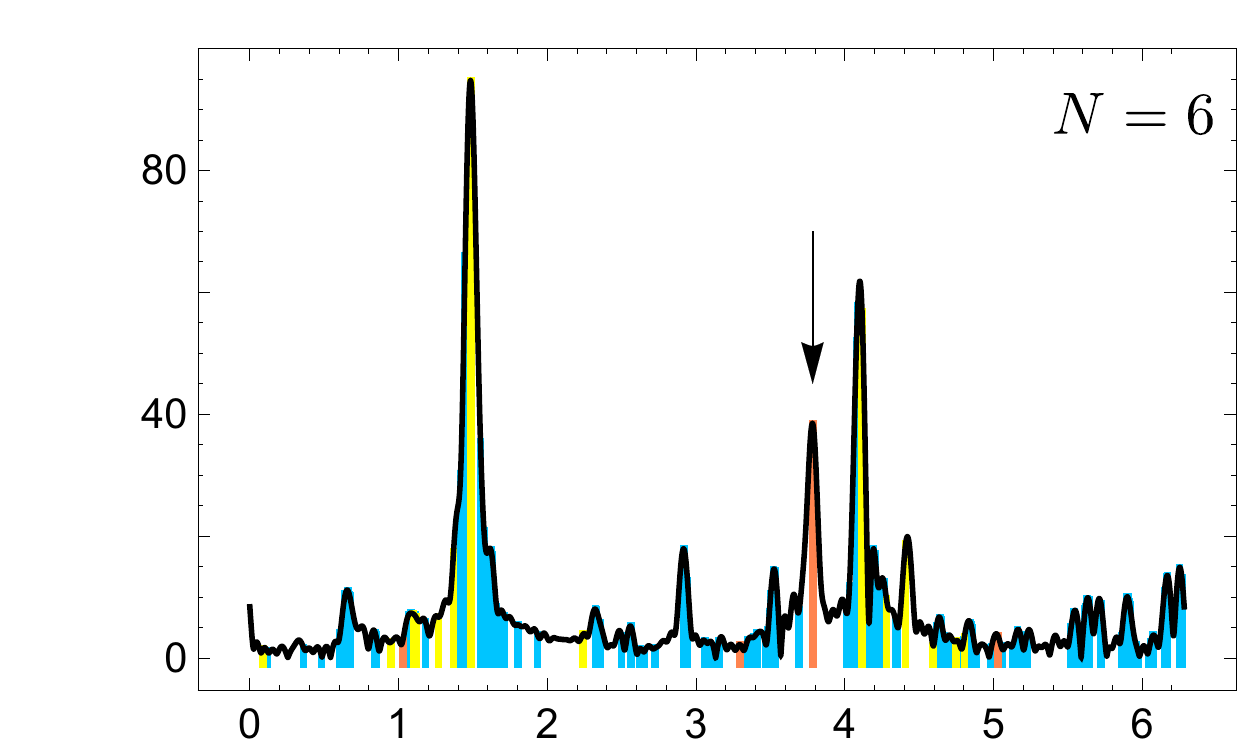}
\hfill
\includegraphics[width=0.32\textwidth]{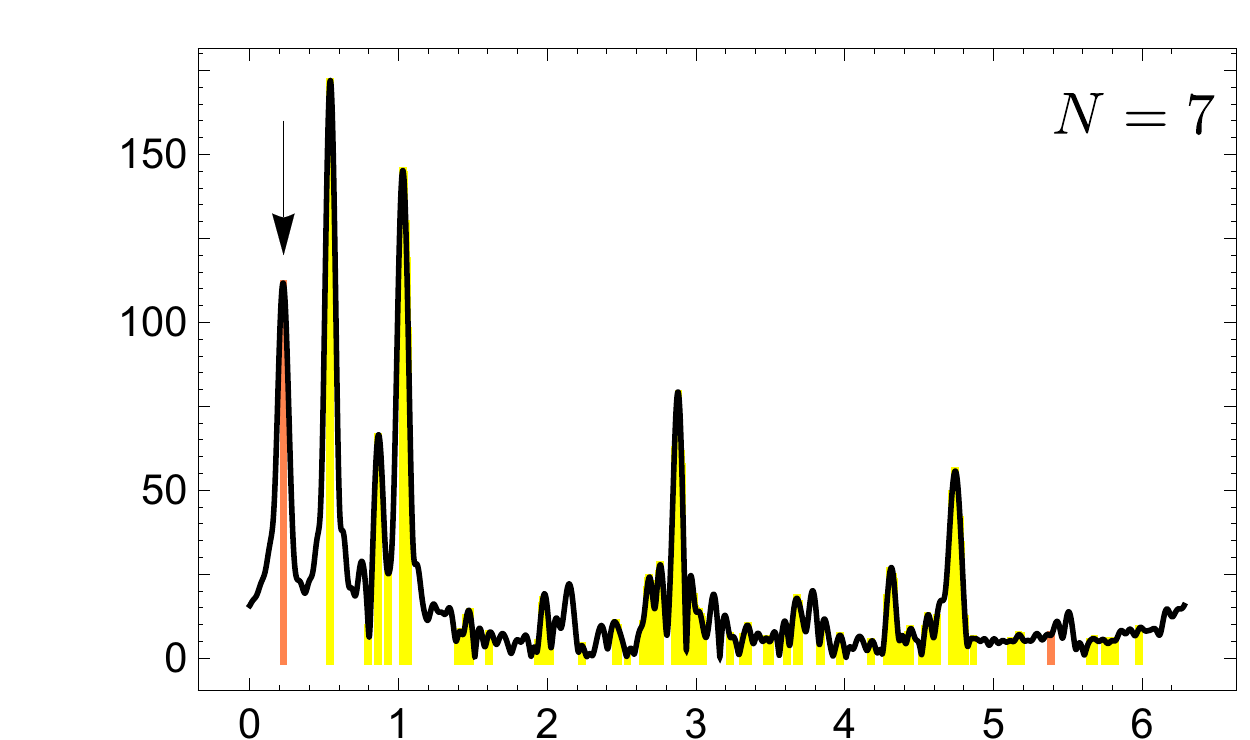}
\\
\hfill
\includegraphics[width=0.32\textwidth]{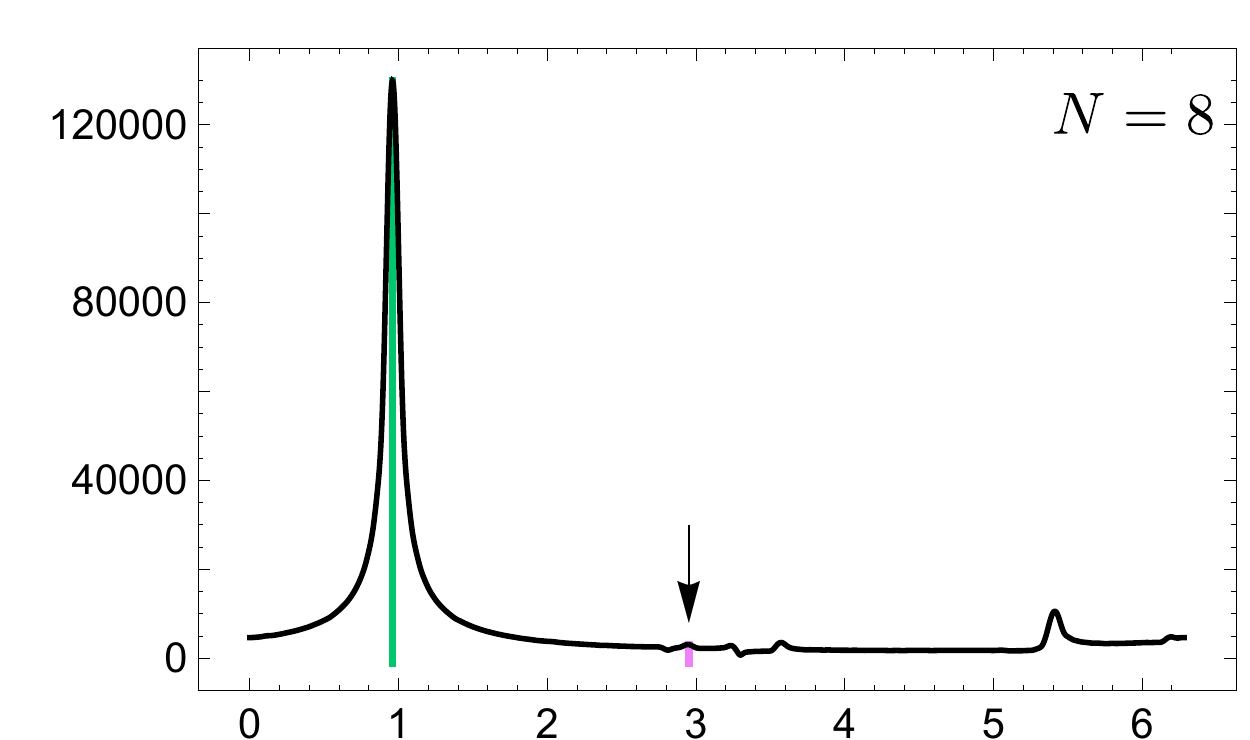}
\hfill
\includegraphics[width=0.32\textwidth]{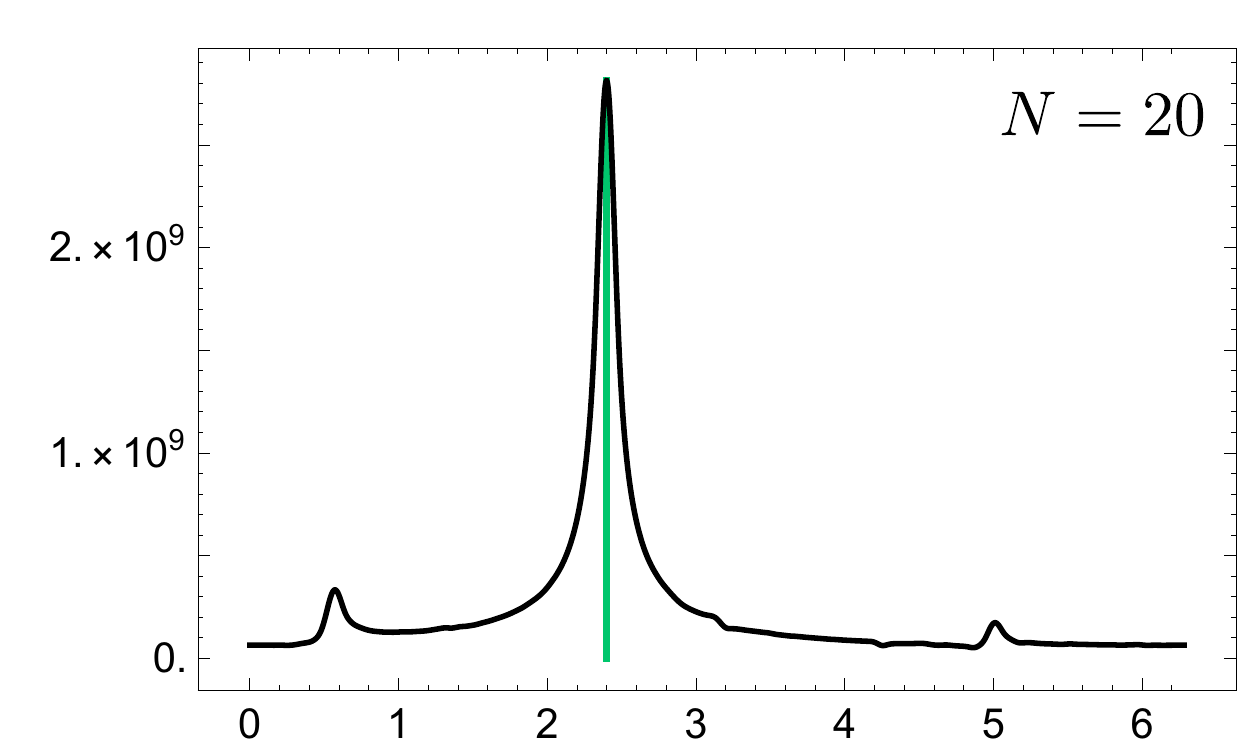}
\hfill
\includegraphics[width=0.32\textwidth]{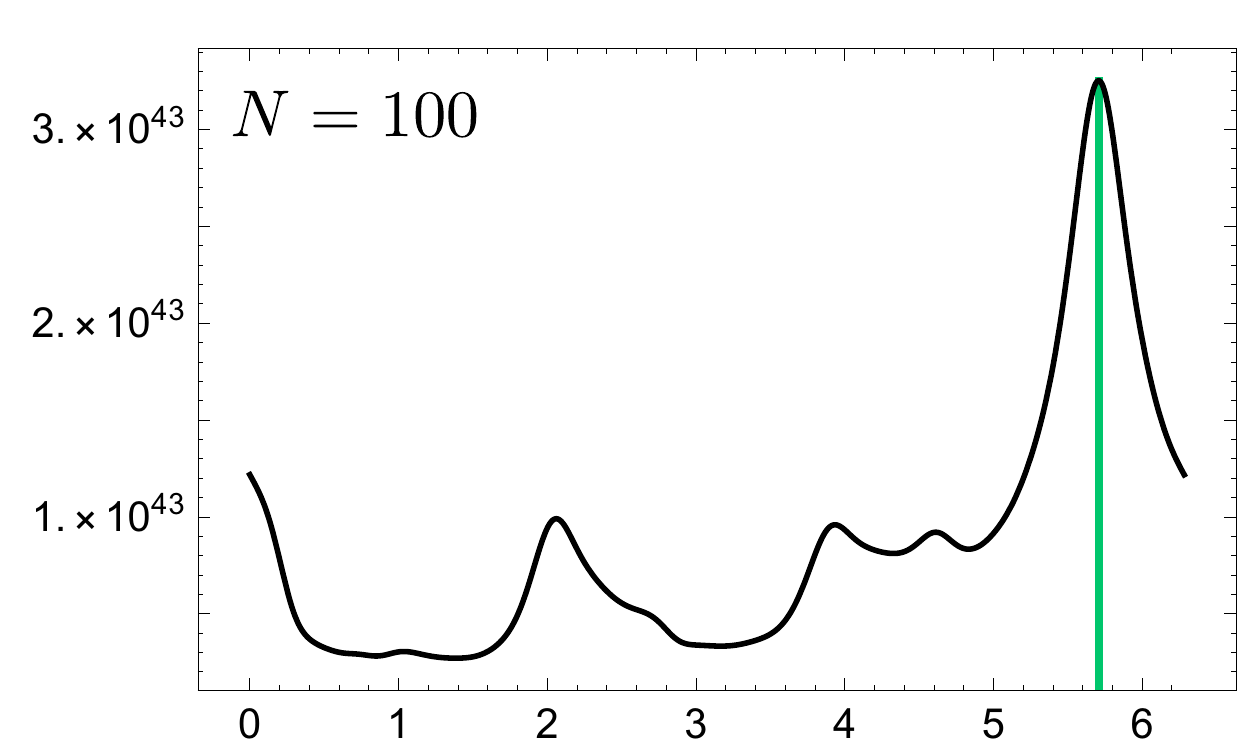}
\caption{For \(T\teq 2\), $|\rho(S)|$ as black curve with
  parameters \(J\teq 0.7\), \(b^x\teq b^z \teq
  0.9\) and \(\jcut\teq 114\). The
  peaks for \(\Nprim\teq N\) are blue, for \(\Nprim\teq 1\)
  red and yellow otherwise. For \(N\teq 4,8,20,100\), very large
  peaks, marked green, due to collective motion occur at actions $S_{\rm man}$.}
\label{fig:sFTT2}
\end{figure*}

\textit{Dominance of collectivity} --- In both spectra in Fig.\ \ref{fig:sFTT1}, collective orbits with
\(\Nprim<N\) play only a minor role. This appears to be a systematic
effect for \(T\teq 1\).   In Fig.~\ref{fig:sFTT2}
we display $|\rho(S)|$ for \(T\teq 2\). We find qualitatively similar pictures for larger $T$.  The most striking features are the
gigantic peaks, arising whenever the particle number is an integer
multiple of four, \(N\teq 4k\). They overshadow all other
structures,
 see the panels for \(N\teq4,\,8,\,20,\,100\). This  is especially remarkable for \(N\teq 100\) given the enormous number of POs. Furthermore, the  scaling exponent $\alpha$ of the peak heights 
 grows linearly with $N$, \(\alpha(N)\sim \alpha_1N\), where $\alpha_1\approx1/5$ in a wide parameter range.
This structure is thus important for large  $N$.
Even \(\alpha(4)\) is significantly larger than the corresponding values for bifurcations in single particle systems.
This clearly indicates
collective motion and, importantly, we can explain it  by the emergence of  certain classical   structures in the phase space of the system as explained below.
In the other panels of Fig.~\ref{fig:sFTT2}, where  \(N\neq 4k\), similar peaks
are not seen.  Due to a large number of POs  we
employ 
for \(N\teq 6,7\) the  filtering condition \(\left|\det(M_\gamma-\mathds{1})\right|<5\cdot 10^3\). The collective orbit highlighted by the arrow, featuring \(\Nprim\teq1\), is
the same for all panels and creates a discernible peak still for
\(N\teq 7\). The reason for this is that it is not sufficiently
isolated but close to a pitchfork bifurcation.

\begin{figure}
\includegraphics[width=0.95\columnwidth]{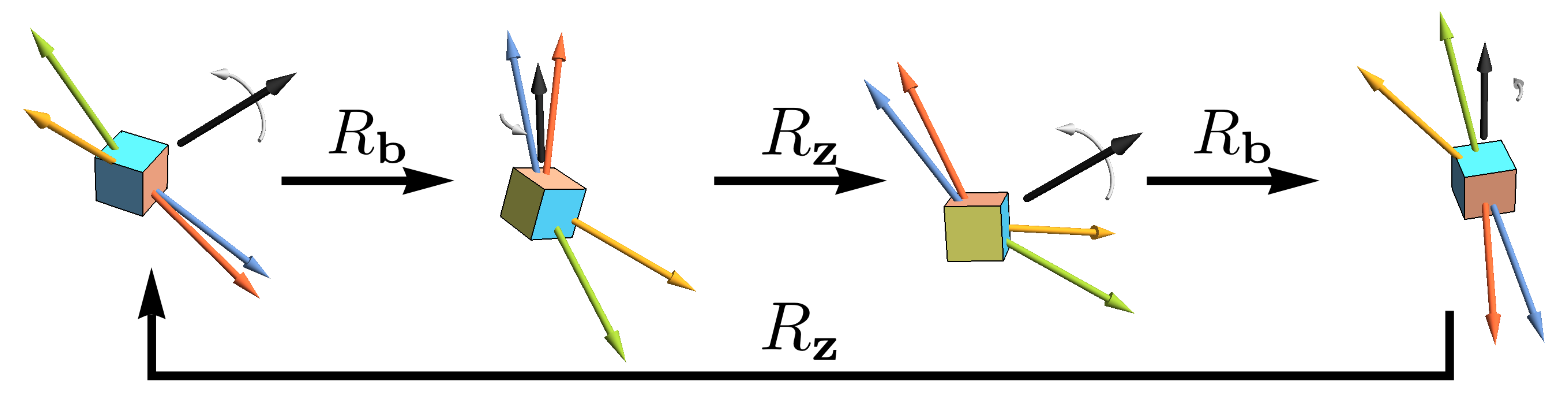}
\caption{Trajectory of a PO on the manifold, depicted
  after each rotation step for \(J\teq 0.7\) and
  \(b^x\teq b^z\teq 0.9\). Spins are ordered according to blue,
  yellow, green and red.}
\label{fig:manifoldSchema}
\end{figure}

\textit{Semiclassical interpretation for the multiple-of-four
  collectivity} --- The  large peaks for \(N\teq 4k\) do not result from isolated orbits, but from  a four-dimensional manifold of POs. To explain this, we return to Eq.\ \eqref{eq:ClassRot}. 
We demand that the Ising rotation angle \(4J\chi_m\) is identical for all spins, $\chi_m=\chi$, such that \(R\teq R_{\bf z} \big( 4J \chi\big)
R_{\bf b}\big(2|\mathbf{b}|\big) \) is a rotation by \(\pi\) around the same axis for
both time steps.
This condition can be met iff $N=4k$. For $N=4$ it imposes four restrictions (two per time step) on the angle of the Ising rotation of each pair of even and odd spins. After elementary calculations we obtain
that the orientation of the spins must satisfy
\begin{eqnarray}\label{cond1}
\chi&=&n_{i-1}^z+n_{i+1}^z\\
\chi&=&\left(n_{i-1}^x+n_{i+1}^x\right) \cot{\beta}
+\left(n_{i-1}^y+n_{i+1}^y\right) \frac{\cot{|\textbf{b}|}}{\sin{\beta}}\nonumber
\,,
\end{eqnarray}
where $\beta\teq\arctan{b^x/b^z}$ is the angle between the magnetic field and the $z$ axis. The angle 
$\chi$ is calculated from
\begin{equation}
b^z\,\tan{\left(2J \chi\right)}=|\textbf{b}|\,\cot{|\textbf{b}|} \,.
\label{eq:ceq}
\end{equation}
The conditions (\ref{cond1}) and (\ref{eq:ceq}) define a four-dimensional manifold. All points of it 
are POs with period $T=2$ due to $R^2=\mathds{1}$ with the action
$S_\text{man}=2 J N \chi^2$.
This manifold exists as repetition also for $N=4k$. 
As \(\chi\) is independent of \(n\) for all spins their motion is
highly correlated. They perform a collective solid body rotation keeping the angles between the spins constant over
time. Fig.~\ref{fig:manifoldSchema} shows a trajectory on
the manifold after the action of each rotation.
In the cases where Eq.~\eqref{eq:ceq} allows more than one possible
solution for \(\chi\), the system behavior is more complicated.
Several manifolds with possible mixtures of the \(\chi_n\) occur, and
their number grows with \(N\). As a result, in this regime \(|\rho(S)|\) attains 
a more complicated profile with several maxima \cite{AkilaI}.

\begin{figure}
\includegraphics[width=0.9\columnwidth]{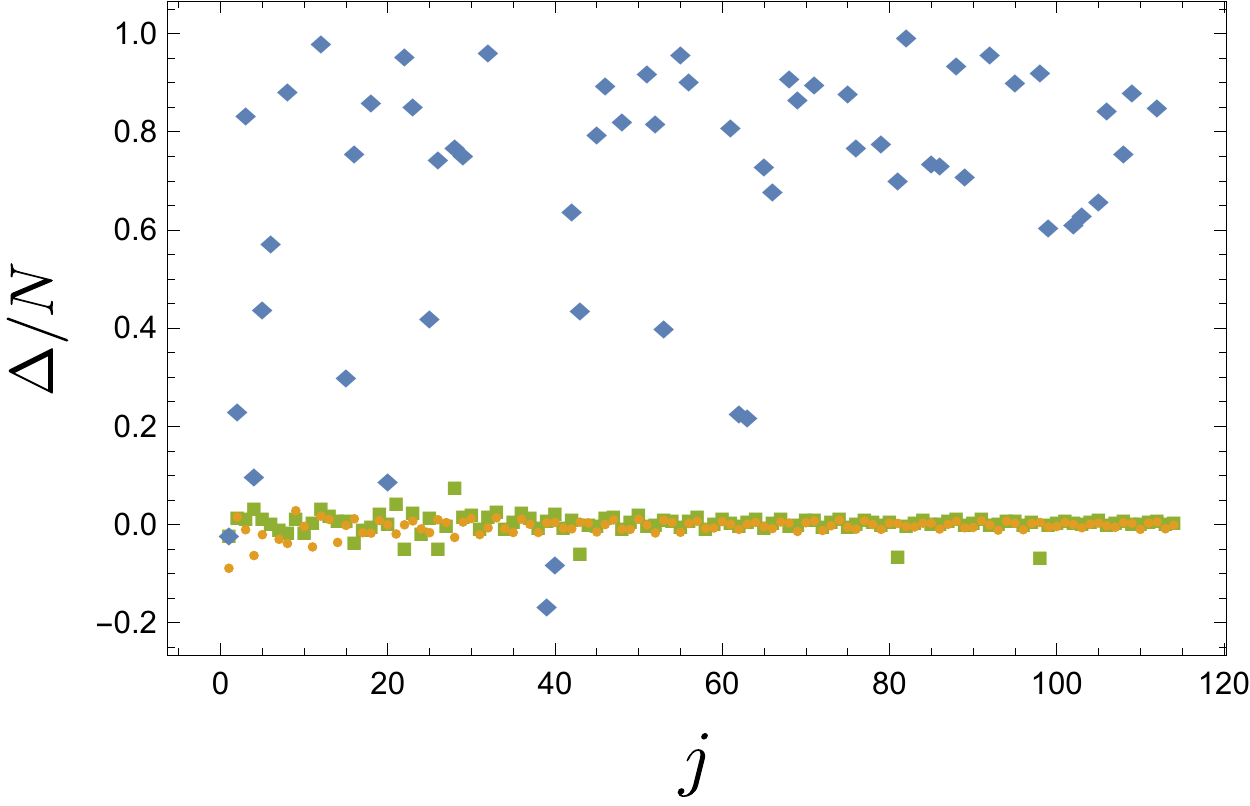}
\caption{Difference \(\Delta(j)\)
  divided by \(N\), for the manifolds with
  $N=4$ as orange circles and with $N=80$ as green squares, and for \(N\teq 3\) as blue diamonds.}
\label{fig:deltaS}
\end{figure}

To quantify the impact  of the PO manifold for fixed $j$ on the quantum spectrum it is instructive to 
study to what extent the phase  of $\Tr \hat{U}^T$ is determined by the action $S_{\rm max}$ of the orbit leading to the largest peak in $|\rho(S)|$. Therefore we introduce 
\begin{equation}
\Delta(j) = \operatorname{Im}\operatorname{Log}\Tr \hat{U}^T-(2j+1)S_{\rm max} \mod 2\pi. 
\label{eq:deltaS}
\end{equation} 
As shown in Fig.\ \ref{fig:deltaS}, for  $N\neq 4k$, $\Delta(j)$ is an wildly fluctuating function of $j$. However, for $T=2$,  $N= 4k$,  $\Delta(j)$ is approximately constant. This implies that the phase of ${\rm Tr} {\hat U}^T$ 
is strongly dominated by the  single term provided by the PO manifold. 
This is due to the structure of the eigenvalues of the dual operator. For large $N$   the duality relationship (\ref{dual}) guarantees that ${\rm Tr} {\hat U}^T$ is dominated  by the eigenvalues of the dual operator \(\widetilde{U}\) with largest magnitude.
We find the remarkable property \cite{AkilaI} that $\widetilde{U}$ possesses four eigenvalues with largest magnitude $a_l{\rm e}^{i\varphi_l}$, $l=1,\ldots,4$ where $a_l= {\rm const.} j^{\alpha_1}(1+O(1/j))$ and
\begin{equation}
\varphi_l = (j+1/2) S_\text{man}/N  +\frac{\pi l}{2}+O(1/j) \mod 2\pi\ .
\label{eq:BS}
\end{equation}
This is formally similar to a Bohr-Sommerfeld quantization rule for $ S_\text{man}$. 
The existence of collective dynamics for $N=4k$ finds here its correspondence in the fact that the contribution  of the four eigenvalues \((a_l)^N \eu^{\iu N \varphi_l}\) to 
$\Tr\hat{U}^T$  cancels due to the $l$-dependence of the phases in (\ref{eq:BS}) except for  $N = 4k$ where $ N \varphi_l$ is independent of $l$. 

\textit{Conclusions} --- We carried out a semiclassical analysis of
a (non-integrable) many-body quantum system. We studied a kicked
spin chain as a representative of a class of systems which presently
is in the focus of experimental and theoretical research. For the first
time, we presented a unifying semiclassical approach to incoherent and to coherent,
collective dynamics. The key tool was a recently discovered duality
relation between the evolutions in time and particle number. It
outmaneuvers the drastically increasing complexity of the problem with
growing particle number. In the spin chain a certain type of
collective motion strongly contributes to the spectra, whenever the
particle number is an integer multiple of four. An experimental
verification is likely to be feasible in view of the improving ability
to control systems with larger numbers of spins.

\clearpage

\begin{widetext}
 \begin{center}
  {\large\bf Supplemental material for ``Semiclassical Identification of Periodic Orbits in a Quantum Many-Body System''}
 \end{center}
\end{widetext}
\textit{Explicit Form of the Dual Operator} ---
In terms of the \((2j+1)^T\) dimensional product basis,
\begin{equation}
|\bm \sigma\rangle= |\sigma_1\rangle \otimes|\sigma_2\rangle\otimes\dots \otimes|\sigma_T\rangle\,
\end{equation}
with discrete single spin states \(\sigma_t \in \{-j,\,\allowbreak-j+1\,\allowbreak,\ldots\, \allowbreak+j \}\), we can provide the explicit form of  the dual operator \(\widetilde{U}\teq \widetilde{U}_I \widetilde{U}_K\).
The interaction part is a diagonal matrix with elements
\begin{equation}
\langle{\bm \sigma}| \widetilde{U}_I | {\bm\sigma'}\rangle
=\delta_{\bm \sigma, \bm{ \sigma'}} \, \prod_{t=1}^T
\langle \sigma_{t}| \exp\left(2\iu \, \boldsymbol{b}\cdot \boldsymbol{\hat{s}}\right)\, | \sigma_{t+1} \rangle\,.
\end{equation}
The boundary conditions are periodic, \ie\ \(\sigma_{T+1}\teq \sigma_1\). The kick part features a local structure
\begin{equation}
\widetilde{U}_K = 
\bigotimes_{t=1}^T \tilde{u}_K\,,
\quad
\langle \sigma | \tilde{u}_K | \sigma' \rangle
=\exp{\frac{4\iu J \sigma \sigma'}{j+1/2}}\,.
\end{equation}
Although \(\tilde{u}_K\) is related to the interaction of \(\hat{U}_I\) it is not diagonal.

An explicit example can be given in the integrable case (\(b^x\teq 0\)) where the components of the dual operator are
\begin{equation}
\begin{split}
\widetilde{U}_{nm}=
\exp \bigg( & \iu \frac{4JT}{j+1/2}(n-j-1)(m-j-1)
\\
& +  2\iu T b^z (n-j-1) \bigg)\,.
\end{split}
\end{equation}
The indices \(m, n\) run from 1 to \(2j+1\) and time turns, in this case only, to a scaling of the system parameters.

\end{document}